\documentclass[12pt, fleqn]{article}
\usepackage{graphicx}
\usepackage{amssymb}

\newcommand{\bra}[1]{\langle#1}
\newcommand{\ket}[1]{|#1\rangle}
\newcommand{\set}[1]{\left\{#1\right\}}
\newcommand{\eps}{\varepsilon}

\begin{document}
\title{Unpredictability of wave function's evolution in nonintegrable quantum systems}
\author{
I. B. Ivanov \\
Theory Department, Petersburg Nuclear Physics Institute,\\
188300 Gatchina, Russia
}
\date{}
\maketitle {\small{\bf Abstract.} It is shown that evolution of
wave functions in nonintegrable quantum systems is unpredictable
for a long time $T$ because of rapid growth of number of
elementary computational operations $\mathcal O(T)\sim T^\alpha$.
On the other hand, the evolution of wave functions in integrable
systems can be predicted by the fast algorithms  $\mathcal
O(T)\sim (log_2 T)^\beta$ for logarithmically short time and thus
there is an algorithmic "compressibility" of their dynamics. The
difference between integrable and nonintegrable systems in our
approach looks identically for classical and quantum systems.
Therefore the minimal number of bit operations $\mathcal O(T)$
needed to predict a state of system for time interval $T$ can be
used as universal sign of chaos. }

\vspace{1cm}

Chaos as universal phenomenon exists in various systems (for
example, in human society) and from this point of view the general
approach to chaos should not be based on particular properties of
a system. It is well known \cite{LL} that motion of nonintegrable
classical systems has all attributes of chaos: complexity,
unpredictability and randomness. However in nonintegrable quantum
systems the apparent signs of chaos seems to be absent \cite{PTP}
and the main direction of studies in the field of "quantum chaos"
is semiclassical analysis of various quantum "signatures" of
classical chaos \cite{Berry, Eckh}. Nevertheless, taking into
account the fundamental correspondence principle it is reasonble
to suppose that evolution of nonintegrable quantum systems should
be also unpredictable. As a basis tool to analyze predictability
of dynamics we use a number of elementary computational operations
$\mathcal{O}(T)$ needed to determine a state of the system for
time interval $T$.

Let us consider firstly how a number of elementary computational
operations needed for prediction of system's evolution depends on
time $T$ and accuracy $\Delta$ in classical mechanics. It is well
known \cite{LL} that a distance $||\delta x(t)||$ between
initially close phase space points grows with time as
\begin{equation}
\label{dx}
||\delta x(t)||=||\delta x(0)||f(t),
\end{equation}
where $f(t)\sim e^{\lambda t}$ for nonintegrable systems and $f(t)\sim t$
for integrable ones. The exponential growth of inevitable computational errors
leads to unpredictability of long time evolution in nonintegrable classical
systems. To predict a state (trajectory) of the system by the moment $T$ with accuracy $\Delta$
we must make computations with accuracy at least
\[
\delta=\frac{\Delta}{f(T)}
\]
and length of mantissa
\[
n\sim -log_2\delta =log_2 f(T)-log_2 \Delta.
\]
It is reasonable to suppose that an algorithm which generates a trajectory
has a power dependence of number of elementary bit
operations $\mathcal O(n)$ on the length of mantissa (for example,
$\mathcal O(n)\sim n$ for addition  and $\mathcal O(n)\sim n^2$
for multiplication):
\begin{equation}
\label{O}
\mathcal O(T)\sim n^\alpha\sim (log_2 f(T)-log_2 \Delta)^\alpha,
\end{equation}
where $\alpha>1$ is some number. Inserting now $f(t)$ in (\ref{O})
we obtain for nonintegrable systems ($\alpha>1$)
\begin{equation}
\label{HPCh} \mathcal O(T)\sim T^\alpha,
\end{equation}
while for integrable ones
\begin{equation}
\label{HPCr} \mathcal O(T)\sim (log_{2}T)^{\alpha}.
\end{equation}

In the first case a number of operations and hence a computational
time needed for prediction grows as a power of time $T$. Therefore
the prediction for a long time is impossible and evolution is not
algorithmically "compressible". In  regular systems there is a
completely different situation because computational time grows
logaritmically with time $T$ and prediction is always possible
("compressibility" of evolution). It is interesting to note that
algorithmic "uncompressibility" of some numerical sequence is the
main sign of its randomness: such sequence we can only observe.

Let us consider now a time evolution of some state $\psi(q, t)$ in
a stationary quantum system $H(q, p)$
\begin{equation}
\label{sheq}
i\hbar\partial_t\psi (q, t)=H\psi(q, t),
\end{equation}
\[
\psi(q, 0)=\psi_0(q),
\]
Any solution of (\ref{sheq}) at $t=T$ can be represented as
\begin{equation}
\label{summa}
\psi(q, T)=\sum_\mu c_\mu e^{-iE_\mu T/\hbar}\phi_\mu(q),
\end{equation}
where $\set{E_\mu}$ and $\set{\phi_\mu}$ -- are exact eigenenergies and eigenfunctions
of Hamiltonian $H$ and $c_\mu =\bra{\phi_\mu}\ket{\psi_0}$.
So if quantum chaos exists it hides in eq. (\ref{summa}) which is essence of quantum unitary
evolution.
In analogy with classical case it would be reasonable to investigate a time dependence of
two initially close wave functions $\psi_1$ and
$\psi_2$. It is natural to characterize a difference between wave functions by the quantity
\begin{equation}
\label{dev}
||\delta\psi||=||\psi_2-\psi_1||=\sqrt{2(1-Re\bra{\psi_1}\ket{\psi_2})}.
\end{equation}
If at $t=0$ we have $||\delta\psi (0)||<\eps$ then at $t=T$ the
difference (\ref{dev}) is also small $||\delta\psi (T)||<\eps$
because from unitarity it follows that $\bra{\psi_1}\ket{\psi_2}$
does not vary with time. Therefore in quantum systems there is no
divergence of initially close states which is the main sign of
chaos in classical systems.

How then unpredictability of evolution can arise in quantum systems?
To predict a state of the system $\psi(q, T)$ at moment $t=T$ it is necessary
to know all eigenvalues and eigenfunctions in (\ref{summa}) exactly.
Really we use in (\ref{summa}) approximate eigenvalues and eigenfunctions
and therefore by some moment exact wave function and
our prediction will be completely different.
It is obvious that to increase a maximal time of accurate prediction
we must improve accuracy of eigenvalues.
The spectrum of integrable systems is easy computable and
computational resources grow slowly with increasing of accuracy
while for nonintegrable systems considerable efforts are needed to improve the spectrum
and thus to increase the time of correct prediction.
Therefore we can assume that quantum integrable systems differ significantly from nonintegrable
ones in a sense of number of bit operations $\mathcal O(T)$ needed to predict a future.
In the remaining part of this Letter we consider some details of
the suggested mechanism of unpredictability and hence of chaos in quantum unitary evolution.

\begin{figure}
\includegraphics[width=\textwidth]{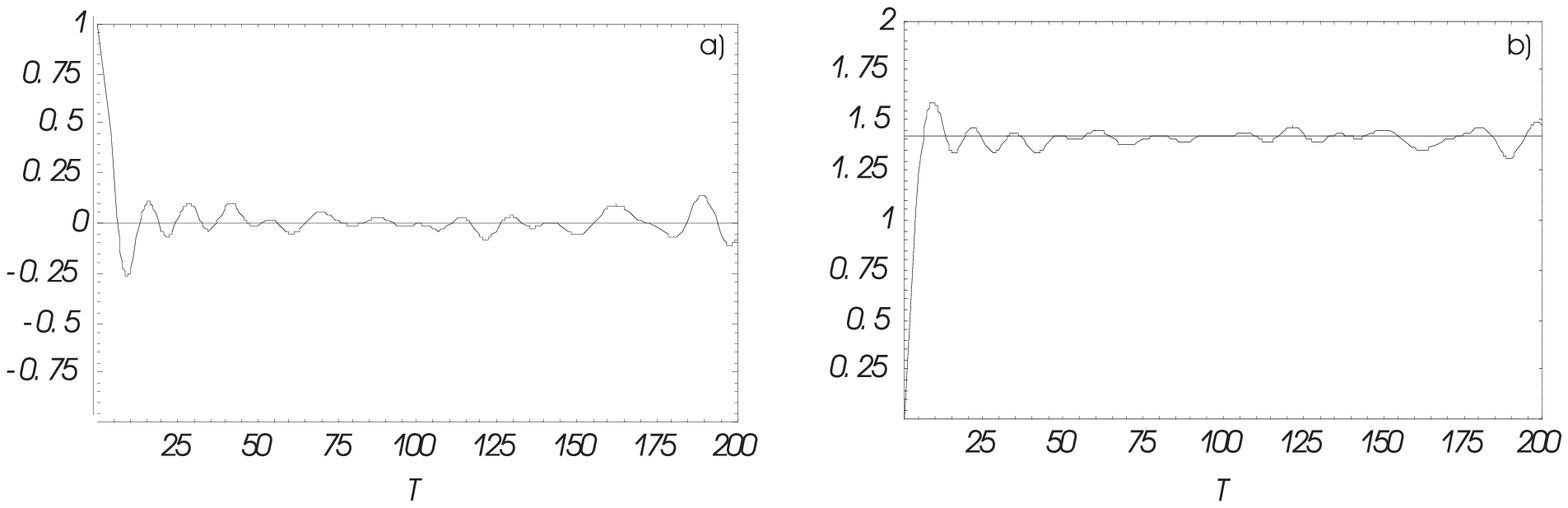}
\caption{A typical behaviour of (a) $\bra{\psi(T)}\ket{\tilde\psi
(T)}$ and (b) $||\delta\psi (T)||$. \label{fig:thp}}
\end{figure}

Let us assume that we have found approximate eigenenergies $\{\tilde E_\mu\}$ and
eigenfunctions $\{\tilde\phi_\mu\}$. Then we can predict by means (\ref{summa})
that at moment $T$ wave function will be
\begin{equation}
\label{asumma}
\tilde\psi (q, T)=\sum_\mu\tilde c_\mu e^{-i\tilde E_\mu T/\hbar}\tilde\phi_\mu(q),
\end{equation}
where $\tilde c_\mu =\bra{\tilde\phi_\mu}\ket{\psi_0}$.
Our main goal now is to find a time dependence of the difference between exact $\psi (q, T)$
and approximate $\tilde\psi (q, T)$ wave functions
\begin{equation}
\label{devT}
||\delta\psi (T)||=||\tilde\psi (T)-\psi (T)||=\sqrt{2(1-Re(\bra{\psi(T)}\ket{\tilde\psi (T)}))}.
\end{equation}
Making use of (\ref{summa}) and (\ref{asumma}) we have:
\begin{equation}
\label{sp}
\bra{\psi(T)}\ket{\tilde\psi (T)}=\sum_{\mu , \nu}c^*_\mu\tilde c_\nu e^{-i(\tilde E_\nu
-E_\mu)T/\hbar}\bra{\phi_\mu}\ket{\tilde\phi_\nu}.
\end{equation}
Now we need some assumptions about accuracy of spectrum obtained.
If a maximal error of eigenfunctions $\tilde\phi_\mu$ does not exceed $\eps$
\[
\delta\phi_\mu =\tilde\phi_\mu -\phi_\mu,
\]
\[
||\delta\phi_\mu|| < \eps.
\]
then the error of $E_\mu$ is
\begin{equation}
\label{dE}
\delta E_\mu =\tilde E_\mu - E_\mu = 2E_\mu Re(\bra{\phi_\mu}\ket{\delta\phi_\mu} )+
\bra{\delta\phi_\mu}|H\ket{\delta\phi_\mu},
\end{equation}
\[
|\delta E_\mu| < 2|E_\mu|\eps + \eps^2||H||,
\]
and error of $c_\mu$ is
\[
\delta c_\mu =\tilde c_\mu -c_\mu =\bra{\delta\phi_\mu}\ket{\psi_0},
\]
\[
|\delta c_\mu | <\eps.
\]
Scalar product $\bra{\phi_\mu}\ket{\tilde\phi_\nu}$ we can write as
\[
\bra{\phi_\mu}\ket{\tilde\phi_\nu}
=\delta_{\mu\nu}+\bra{\phi_\mu}\ket{\delta\phi_\nu}=\delta_{\mu\nu}+R_{\mu\nu},
\]
\[
|R_{\mu\nu}|<\eps.
\]
Now (\ref{sp}) becomes
\begin{equation}
\label{spf}
\bra{\psi(T)}\ket{\tilde\psi (T)}=\sum_{\mu}c^*_\mu\tilde c_\mu (1+R_{\mu\mu})
e^{-i\delta E_\mu T/\hbar}+\sum_{\mu\neq\nu}c^*_\mu\tilde c_\nu R_{\mu\nu}
e^{-i(E_\nu -E_\mu +\delta E_\nu )T/\hbar}.
\end{equation}
It is clear that first sum carries the main contribution in
$\bra{\psi(T)}\ket{\tilde\psi (T)}$ because second sum has a small
multiplier $R_{\mu\nu}$. To understand the main features of
(\ref{sp}) it is sufficient to investigate time dependence of
\begin{equation}
\label{spfa}
\bra{\psi(T)}\ket{\tilde\psi (T)}\approx\sum_{\mu}|c_\mu|^2e^{-i\delta E_\mu T/\hbar}.
\end{equation}
At $T=0$ a value of $\bra{\psi(0)}\ket{\tilde\psi (0)}$ equals $1$, then
$\bra{\psi(T)}\ket{\tilde\psi (T)}$ decreases and at  some moment
$T_p$ reaches  values nearly zero.
After this $\bra{\psi(T)}\ket{\tilde\psi (T)}$ starts to oscillate about zero with
 some small amplitude. For times $T>T_p$ the value of $||\delta\psi (T)||$ oscillates
about mean value $\sqrt{2}$ which corresponds to
completely independent states. This typical behaviour of $\bra{\psi(T)}\ket{\tilde\psi(T)}$
and  $||\delta\psi (T)||$ is shown in Fig. \ref{fig:thp}.
This means that for $T>T_p$ wave function $\tilde\psi (T)$
completely differs from exact function $\psi (T)$ and our further prediction is impossible.

We want now to find how $T_p$ and average amplitude of small
oscillations $\bar A$ depend on errors $\delta E_\mu$ and
coefficients $c_\mu$. Our intuition and numerical experience say
that $T_p$ depends on dispersion of errors $dE$ while $\bar A$ is
a function of number $dim$ of the most important addendums in
(\ref{spfa}). It is obvious that $\bra{\psi(T)}\ket{\tilde\psi
(T)}$ becomes nearly zero when arguments of exponents in
(\ref{spfa}) fill interval $[0, 2 \pi]$. From this condition we
have $T_p$:
\begin{equation}
\label{Tp}
T_p\sim\frac{2 \pi}{2 dE/\hbar} =\frac{\pi \hbar}{dE}.
\end{equation}
To estimate $\bar A$ let us make further simplification that all
$c_\mu$ in (\ref{spfa}) are equal $1/\sqrt{dim}$. In such a case
(\ref{spfa}) becomes
\begin{equation}
\label{Resp}
P(T)=Re(\bra{\psi(T)}\ket{\tilde\psi (T)})=\frac{1}{dim}\sum_{\mu=1}^{dim} \cos (\delta E_\mu T/\hbar )
\end{equation}
and $P^2(T)$ is:
\[
P^2(T)=\frac{1}{dim^2}\sum_{\mu,\nu}\cos (\delta E_\mu T/\hbar )
\cos (\delta E_\nu T/\hbar )=
\]
\[
\frac{1}{dim^2}\sum_{\mu}\cos^2{(\delta E_\mu T/\hbar )}+
\frac{1}{dim^2}\sum_{\mu\neq\nu}\cos (\delta E_\mu T/\hbar )
\cos (\delta E_\nu T/\hbar ).
\]
It is easy to see that if we make time averaging of $P^2(T)$ then
the first sum equals $dim/2$ while the second sum is zero because
arguments of cosines are almost independent for different $E_\mu$
and $E_\nu$. In the result average amplitude of oscillations of
$Re(\bra{\psi(T)}\ket{\tilde\psi (T)})$ can be estimated as
\begin{equation}
\label{Amp}
\bar A\sim\sqrt{\overline{P^2(T)}}=\frac{1}{\sqrt{2 dim}}.
\end{equation}

We tested our theoretical estimates (\ref{Tp}) and (\ref{Amp}) for
wide range of parameters and found good agreement with numerical
computations. So we have obtained a quite reasonable result that
uncertainty in our knowledge of spectrum $E\pm dE$ leads to
limitation in prediction time of $\psi(T)$ evolution
$T_p\sim\hbar/dE$.

The next step consists of estimating of number of operations
needed to obtain the spectrum with some accuracy $dE$. If the
spectrum of a system can be computed by some formulae or by
effective algorithms (as in the case of integrable systems), the
number of operations grows slowly with decreasing of error $dE$.
To explain this let us assume for simplicity that to compute the
spectrum we must perform some series of successive actions which
do not depend on accuracy required (for example $E_N=\hbar\omega
(N+1/2)$). Obviously the necessary number of bit operations
$\mathcal O(dE)$ depends on mantissa length $n\sim -log_2dE$ only
\begin{equation}
\label{Oint}
\mathcal O(dE)\sim n^\gamma \sim (-log_2 dE)^{\gamma},
\end{equation}
where $\gamma>1$ -- some coefficient characterizing the algorithm of computations.
So we have obtained that in integrable systems a number of bit operations
depends logarithmically on accuracy $dE$.

Fortunately the variety of systems in nature is not exhausted by
easy computable and predictable systems. Description of any
nonseparable system with strong coupling and especially of its
excited states is nontrivial problem even for two degrees of
freedom. The only reliable way to calculate such systems is to use
various variation methods, i.e. to minimize the energy functional
over some space of trial functions. Let us consider how the number
of operations depends on spectrum accuracy for variational method
of Ritz. It is well known \cite{Mihl} that Ritz's method has a
power convergence, i.e. approximate eigenfunctions
$\tilde\phi_\mu$ tend to exact ones $\phi_\mu$ as
\begin{equation}
\label{CnvF}
||\tilde\phi_\mu -\phi_\mu||\sim D^{-\alpha},
\end{equation}
where $D$ is dimension of space of trial functions and $\alpha\sim
1$ -- some coefficient. Obviously (see (\ref{dE})) the rate of
convergence of eigenvalues does not exceed the power law:
\begin{equation}
\label{CnvE}
|\tilde E_\mu -E_\mu|\sim D^{-\alpha}.
\end{equation}
To make use of Ritz's method it is necessary to calculate firstly
a Hamiltonian matrix with number of elements $D^2$, and then to
find its eigenvalues. The total number of operations to do this
can be estimated as $\mathcal O(D)\sim D^\beta, \beta >2$. To
achieve an accuracy $\eps$ the basis dimension $D\sim
\eps^{-1/\alpha}$ is required (see (\ref{CnvE})) and hence the
total number of operations is $\mathcal O(\eps)\sim
\eps^{-\beta/\alpha}$. Keeping in mind the relation between time
of prediction and accuracy of eigenvalues (\ref{Tp}) we obtain the
number of bit operations which are needed to predict the wave
function's evolution over time interval $T$:
\begin{equation}
\label{HPQh}
\mathcal O(T)\sim T^{\beta/\alpha}.
\end{equation}
For regular systems (\ref{Oint}) the number of operations is equal
to
\begin{equation}
\label{HPQr}
\mathcal O(T)\sim (log_2 T)^{\gamma}.
\end{equation}
It should be noted that our estimate (\ref{HPQh}) is valid for any
algorithm with power convergence.

So we see that difference between integrable and nonintegrable
systems in our approach looks identically in classical and quantum
systems. The evolution of integrable systems can be predicted by
fast algorithms for logarithmically short time ("compressibility"
of evolution) while in nonintegrable systems such
"compressibility" is absent. The reason of unpredictability is
universal in classical and quantum systems --- rapid growth of
number of elementary computational operations needed for
prediction, but the mechanism of chaos is completely different.

\end{document}